\documentclass[aps,groupedaddress,twocolumn,10pt]{revtex4}
\usepackage[dvips]{graphicx}
\usepackage{amssymb}

\newcommand{\ped}[1]{\ensuremath{_{\rm #1}}}
\newcommand{\unit}[1]{\ensuremath{{\rm\,#1}}}

\begin{document}

\title{Temperature and junction-type dependency of Andreev reflection in MgB$_{2}$}

\author{R.\@S. Gonnelli} \email[Corresponding
author. E-mail:]{gonnelli@polito.it} \affiliation{INFM -
Dipartimento di Fisica, Politecnico di Torino, 10129 Torino,
Italy}
\author{A. Calzolari}
\affiliation{INFM - Dipartimento di Fisica, Politecnico di
Torino, 10129 Torino, Italy}
\author{D. Daghero}
\affiliation{INFM - Dipartimento di Fisica, Politecnico di
Torino, 10129 Torino, Italy}
%
%
\author{G.\@A. Ummarino}
\affiliation{INFM - Dipartimento di Fisica, Politecnico di
Torino, 10129 Torino, Italy}
\author{V.\@A. Stepanov}
\affiliation{P.N. Lebedev Physical Institute, Russian Academy of
Sciences, Moscow, 117924 Russia}
\author{P. Fino}
\affiliation{Dipartimento di Scienza dei Materiali ed Ing.
Chimica, Politecnico di Torino, 10129 Torino, Italy}
\author{G. Giunchi}
\affiliation{EDISON S.p.A., Foro Buonaparte 31, 20121 Milano,
Italy}
\author{S. Ceresara}
\affiliation{EDISON S.p.A., Foro Buonaparte 31, 20121 Milano,
Italy}
\author{G. Ripamonti}
\affiliation{EDISON S.p.A., Foro Buonaparte 31, 20121 Milano,
Italy}
\date{\today}

\begin{abstract}
We studied the voltage and temperature dependency of the dynamic
conductance of normal metal-MgB$\ped{2}$ junctions obtained either
with the point-contact technique (with Au and Pt tips) or by
making Ag-paint spots on the surface of MgB$\ped{2}$ samples. The
fit of the conductance curves with the generalized BTK model gives
evidence of pure \emph{s}-wave gap symmetry. The temperature
dependency of the gap, measured in Ag-paint junctions (dirty
limit), follows the standard BCS curve with $2
\Delta/k\ped{B}T\ped{c}=3.3$. In out-of-plane, high-pressure point
contacts we obtained almost ideal Andreev reflection
characteristics showing a single small \emph{s}-wave gap $\Delta=
2.6 \pm 0.2$ (clean limit).\\ \\
\textit{Keywords}: A. intermetallic compounds, A. superconductors,
D. superconductivity
\end{abstract}
\maketitle

An enormous interest has been aroused by the recent discovery of
superconductivity in magnesium diboride by Nagamatsu \emph{et al.}
\cite{ref1}. The critical temperature ($T\ped{c}$) of this
intermetallic compound is about 40~\unit{K}. Many papers have
recently appeared in literature concerning the determination of
the energy gap $\Delta$ from tunneling
\cite{ref2,ref3,ref4,ref5,ref6}, Andreev reflection
\cite{ref6,ref7,ref8} and many other measurements. The results are
still controversial, since the measured values of $\Delta$ range
between 2 and 7.5~\unit{meV}. Many authors found a temperature
dependency of the gap following rather well the BCS curve for a
pure \emph{s}-wave superconductor \cite{ref5,ref7,ref8}. On the
other hand, very recent tunnel and Andreev reflection measurements
have suggested the presence in MgB$_2$ of two different gaps, both
having a BCS-like temperature dependency and the same
$T_\mathrm{c}$ \cite{ref20,ref21}. According to a recent
theoretical work \cite{ref22}, the small gap $\Delta_\mathrm{1}$
would appear on the sheets of the Fermi surface arising from 3D
$p\ped{z}$ bonding and antibonding bands, while the larger one
$\Delta_\mathrm{2}$ would be related to the nearly-cylindrical
hole sheets around the $\Gamma$-A line arising from quasi-2D
$p\ped{x,y}$ B bands.

At the present moment, it is of primary importance to understand
whether these two gaps are really present (e.g., by possibly
measuring each of them separately and independently of the other)
and to try to explain why various experiments have so far observed
such a large spread of gap values. To do so, in this paper we
present and discuss Andreev reflection results obtained in normal
metal -- MgB$\ped{2}$ junctions made either with the point-contact
technique (by using sharp platinum or gold tips) or by making
small silver-paint spots on the surface of the samples. We will
show that the fit of the normalized experimental conductance
curves of the junctions with the Blonder-Tinkham-Klapwijk (BTK)
model generalized by Y.~Tanaka and S.~Kashiwaya \cite{ref14} gives
evidence for a \emph{s}-wave symmetry of the order parameter and
shows that, when the tip is almost perpendicular to the B planes,
a single small gap is measured. The results of the fit also
suggest that the superconducting properties of this compound are
strongly influenced by impurities.

The polycrystalline MgB$\ped{2}$ starting material, of high
density (2.4 gr/cm$^3$), was produced at EDISON S.p.A. by means of
a reaction sintering, for 3 hours at 950~\unit{^{\circ} C}, of
elemental B and Mg, in a sealed stainless steel container, lined
with a Nb foil. More details on the preparation are given
elsewhere \cite{ref15}. X-ray powder diffraction showed the
presence in the material of spurious phases (mainly MgO and
unreacted Mg). Both AC susceptibility and resistivity measurements
indicate that $T\ped{c} = 38.8$~\unit{K} with $\Delta T\ped{c} =
0.5$~\unit{K} as determined by the full width at half maximum of
the $d\chi^{\prime}/dT$, where $\chi^{\prime}$ is the real part of
the susceptibility \cite{ref13}. These values, together with the
rather small residual resistivity ($\rho_{0} \approx 4
\mu\Omega\cdot$cm) prove that the high quality of the bulk
material is not spoilt by the presence of spurious phases.
\begin{figure}[h]
\begin{center}
\includegraphics[keepaspectratio,width=7cm]{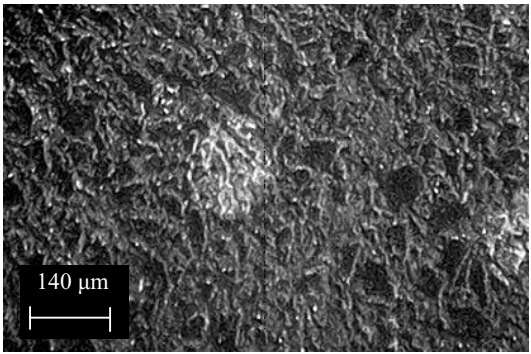}
\vspace{-4mm}\caption{\small{ Photograph of the polished sample
surface, taken by an optical microscope at 98$\times$. Darker
areas are the mirror-like surfaces of single crystals. The bright
spot in the middle is the trace of a contact obtained with
Ag paint.}}\label{f:crystals}
\end{center}
\end{figure}
The samples, of about $2 \times 1 \times 0.5$ mm$^3$, were cut
from the very hard bulk MgB$_{2}$ material with a fine diamond
circular saw. Their surface was polished with fine diamond tools
and then etched with a solution of 1\% HCl in pure ethanol. The
resulting samples, observed by a metallographic microscope,
clearly show the presence of small (50-70 $\mu$m) single crystals
(with random orientation) immersed in a more amorphous matrix.
Figure~1 shows a photograph of a sample surface, taken by an
optical microscope at 98$\times$, where the MgB$_2$ crystals are
clearly visible.

The point-contact junctions were obtained by pressing very sharp
metallic tips against the surface of the samples. We used three
kinds of tips, obtained by electro-chemically etching in a
HCl-HNO$_3$ solution thin Au and Pt wires (with diameter
$\varnothing = 0.2$~mm) and of a thicker Pt wire ($\varnothing =
0.5$~mm), respectively. As a consequence, we could apply different
pressures in the contact area of the sample surface.

Junctions of a different kind were obtained by directly gluing
thin ($\varnothing =25\mu$m) Au wires on the etched surface of the
sample by using small spots of silver paint \cite{ref8}. The
conductance vs. voltage curves showed that many of these contacts
were actually S--N junctions.

We measured the $I-V$ characteristics of the various S--N
junctions and calculated the dynamical conductance curves
(d$I$/d$V$ vs. $V$), which were then normalized so that d$I$/d$V
\simeq 1$ for $|eV| \gtrsim 15$ meV. In doing this we considered
only the data sets for which d$I$/d$V$ was reasonably constant at
$|V|
> 15$ mV and did not show sensible variations at the change of
temperature.

None of the conductance curves showed effects of heating
phenomena. In fact, the values of the normal-state point-contact
resistance we obtained clearly indicate that all the junctions are
in the Sharvin limit (mean free path larger than the size of point
contact)\cite{Sharvin}. The contact radius $a$, evaluated from the
contact resistance, is always less than 100~$\mathrm{\AA}$. Since
the mean free path for MgB$\ped{2}$ is estimated in $600$~\AA~
\cite{ref18}, the conditions for energy-resolved spectroscopy are
totally fulfilled.

\begin{figure}[t]
\begin{center}
\includegraphics[keepaspectratio,width=7cm]{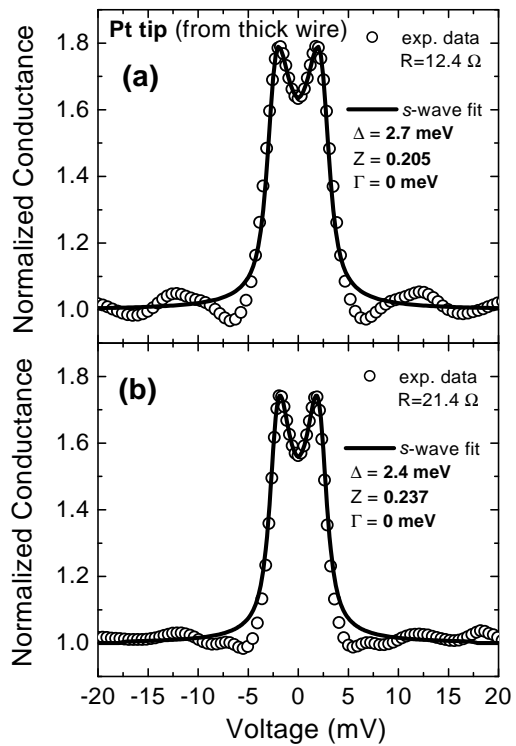}
\vspace{-3mm}\caption{\small{Two examples of experimental Andreev
reflection curves (open circles) obtained at $T=4.2$~K with a tip
made from thick ($\varnothing = 0.5$~mm) platinum wire. The solid
lines are the best-fit curves given by the generalized BTK model
in $s$-wave symmetry.}}\label{f:condEdison}
\end{center}
\vspace{-3mm}
\end{figure}

Figure~2 reports two examples of the best-quality normalized
conductance curves (open circles) obtained at $T=4.2$~K with the
point-contact technique by using a Pt tip made starting from the
thicker Pt wire. A maximum pressure of the order of 0.6 GPa is
applied to the sample by this kind of tips, as we independently
measured from the tip deformation. The conductance curves show
classical Andreev reflection features.  The maximum value of the
normalized conductance is very high in comparison with previous
data present in literature \cite{ref7,ref8,ref21}, being equal to
the theoretical one for an ideal S--N junction with a very low
barrier height according to the BTK model \cite{ref14}. Notice
that contacts of such a high quality were actually obtained in a
small percentage of measurements. We fitted these conductance
curves by using the generalized BTK model \cite{ref14}, and we
found that the best results (see for example the solid lines in
Fig.~2) were obtained with a pure $s$-wave gap symmetry. Let us
remind that the free parameters of an $s$-wave fit are: the
parameter \emph{Z},  which takes into account the barrier height
and the mismatch between the Fermi velocities in the
superconductor and in the normal metal, the lifetime broadening
$\Gamma$, and the value of the gap $\Delta$. Actually, curves such
as those reported in Fig.~2 can be fitted very well by using null
or very small values of $\Gamma$. The resulting gap values are
consistent with one another and give an average value $\Delta =
2.6 \pm 0.2$ meV \cite{ref8,ref21}.
\begin{figure}[t]
\begin{center}
\includegraphics[keepaspectratio,width=7cm]{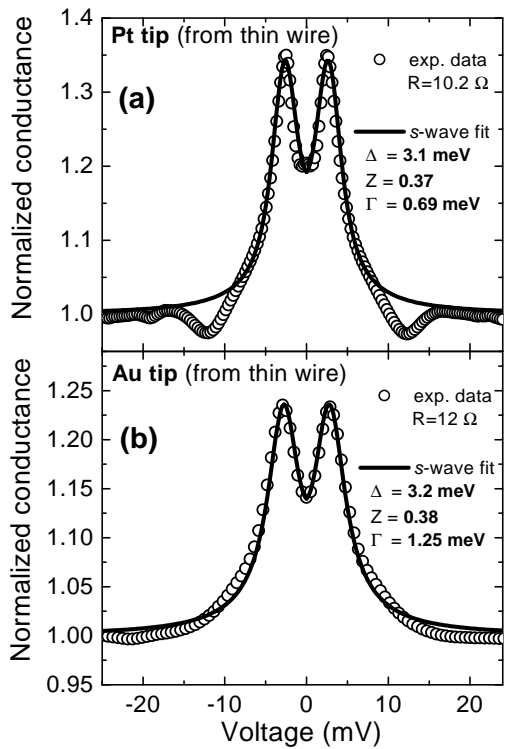}
\end{center}
\vspace{-3mm}\caption{\small{Experimental normalized conductance
curves (open circles) obtained at $T=4.2$~K with platinum (a) and
gold (b) tips made from the thinner wires ($\varnothing=0.2$~mm).
Solid
lines represent the $s$-wave best-fit curves.}}\label{f:condEdison1}
\vspace{-3mm}
\end{figure}

Figure~3 shows two examples of the best point-contact normalized
conductance curves (open circles) obtained with Pt (a) and Au (b)
tips made by starting from the thinner wires
($\varnothing=0.2$~mm), together with the relevant $s$-wave best
fit curves (solid lines). Here, the maximum pressure applied by
the tips on the sample surface is about 0.4 and 0.1 GPa for Pt and
Au tips, respectively.

It can be clearly seen by comparing Fig.~3 (a) and (b) to Fig.~2
that, at the decrease of the tip pressure, the maximum value of
the normalized conductance decreases from $\sim 1.8$ (Pt tip made
from thick wire) to $\sim 1.35$ (Pt tip, thin wire) and $\sim
1.25$ (Au tip, thin wire). The corresponding best-fit parameters
(shown in the legends) indicate a slight increase of $Z$ and
$\Delta$ (up to $Z \sim 0.38$ and $\Delta \sim 3.2$~meV,
respectively) but especially a more remarkable increase of the
lifetime broadening (up to $\Gamma = 0.69$ and $1.25$~meV for Pt
and Au tip, respectively) which seems to indicate a progressive
increase of the disorder in the junction at the decrease of the
tip pressure.

We also obtained good and reproducible Andreev reflection
characteristics in the MgB$\ped{2}$/Ag-spot junctions. The great
stability of these contacts allowed us to study the temperature
dependency of the conductance curves. In figure~4 (a) we report
the normalized conductance curves obtained at different
temperatures between 4.2 K and the temperature at which the
Andreev features disappear, $T_\mathrm{c}^\mathrm{j} \simeq
34.5$~K (open simbols). For clarity, only some of the measured
curves are shown. It is evident from the comparison of these
curves with those of the previous figures that the Andreev
features are largely broadened and the maximum normalized
conductance at low $T$ is reduced to values of the order of 1.15.
The fit of the curves with the generalized BTK model is very good
for any temperature up to $T_\mathrm{c}^\mathrm{j}<
T_\mathrm{c}^\mathrm{bulk}$ if an \emph{s}-wave gap symmetry is
used. The best-fit curves are reported in figure~4 (a) (solid
lines). The fact that $T_\mathrm{c}^\mathrm{j}$ is smaller than
$T_\mathrm{c}^\mathrm{bulk}$ could be due to the presence of a
modified layer at the surface of the polycrystalline samples.

Figure~4 (b) reports the temperature dependency of the order
parameter $\Delta$ (solid circles) determined by the fit of the
curves in Fig.~4 (a), together with the $\Delta$ vs. $T$ standard
BCS behaviour (solid line) calculated with $T\ped{c}^{j} =
34.5$~\unit{K} and $\Delta = 4.9$~\unit{meV}, from which a ratio
$2\Delta / k_\mathrm{B}T_\mathrm{c}^\mathrm{j} = 3.3$ is obtained.
The agreement between experimental data and theoretical curve is
rather good for $T>12$~\unit{K}. Notice that both the theoretical
BCS low-temperature gap $\Delta = 4.9$~\unit{meV} and the
experimental one (which is slightly lower) reported in Fig.~4 are
very different from the gap determined with the point-contact
technique on the same samples (about $\Delta = 2.8\pm
0.4$~\unit{meV} if all the point-contact results are averaged).
\begin{figure}[t]
\begin{center}
\includegraphics[keepaspectratio,width=7cm]{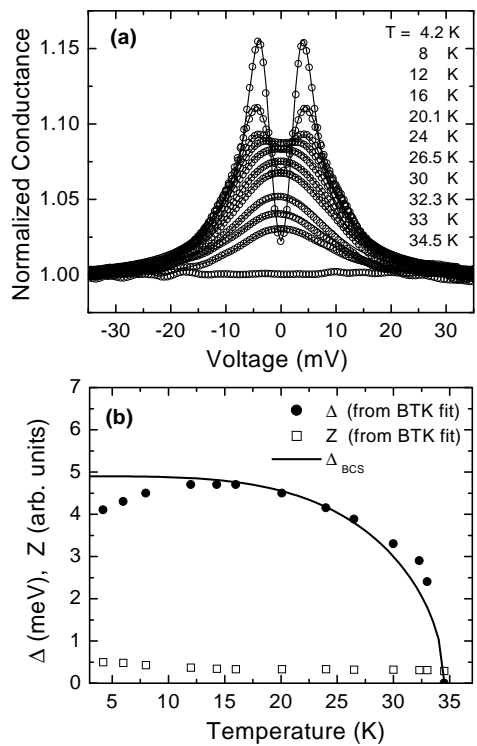}
\vspace{-2mm}\caption{\small{(a) Temperature dependency of the
experimental normalized conductance curves (open circles) for
MgB$\ped{2}$/Ag junctions. Solid lines represent the $s$-wave
best-fit curves obtained with the generalized BTK model. (b)
Temperature dependency of the gap $\Delta$ and of the barrier
parameter $Z$
from the fits of (a).}}\label{f:tempdip}
\end{center}
\vspace{-3mm}
\end{figure}

At a first glance, the results collected in the different sets of
junctions appear inconsistent: very different gap values have been
obtained and a large broadening dominates the Andreev curves in
the Ag-spot junctions, while it is totally absent in the
high-pressure point-contact ones and partially present in the
low-pressure ones. The situation can be clarified by carefully
analyzing the different measurement conditions and the probable
state of the sample surfaces.

In the point-contact junctions the tip touched the sample
perpendicularly to a surface such as that shown in Fig. 1. Due to
the presence of rather large MgB$_2$ crystals (the darker islands
in the figure) and to the small tip dimensions (a few microns), it
is very likely that the best Andreev curves we have shown in Fig.s
2 and 3 are due to contacts with the surface of a single crystal.
Further support to this hypothesis comes from the inspection of
the tip position at the end of the experiments.

The crystal surface, even shortly after the etching, is likely to
be modified by the presence of impurities (e.g. MgO whose
formation is due to the contact with air), grain boundaries or,
maybe, by intrinsic surface phenomena of relaxation or
reconstruction. If the tip applies a rather large pressure on the
sample surface (as in the case of the tips obtained from the thick
Pt wire, see Fig.~2) it is possible to remove or perforate this
modified surface layer. This allows us to obtain ideal contact
with the crystal surface in the clean limit. In this conditions we
have evidence of a \emph{single small gap} $\Delta_\mathrm{1}
\approx 2.6$ meV with a pure \emph{s}-wave symmetry and without
any broadening. The experimental detection of a single small gap
(of about the same amplitude as that we observed) is predicted in
the case of tunneling perpendicular to the honeycomb B planes in a
recent paper by Liu \emph{et al.} \cite{ref22}, where the small
gap $\Delta_1$ and a larger one $\Delta_2 \sim 3 \Delta_1$ are
associated to the 3D parts and to the quasi-2D sheets of the FS,
respectively. The strong directionality of the tunneling
experiments is partly lost in the point-contact measurements,
where the carriers are actually injected in the whole half-space.
Nevertheless, the probability of normal injection is still maximum
and therefore the electronic properties are mainly probed along
the direction perpendicular to the surface. Then, our results
could be compatible with the predictions of Ref.\cite{ref22} if
the tip is within a certain solid angle around the out-of-plane
direction. Taking into account the random distribution of grain
orientation and the small percentage  of conductance curves such
as those reported in Fig. 2, this hypothesis is definitely
plausible.

At the decrease of the tip pressure (see Fig.~3), the modified
surface layer plays a more and more important role giving rise to
a larger $Z$ value but, especially, to a large increase of the
lifetime broadening. If we roughly interpret $\Gamma$ as a measure
of the disorder of the surface, we can argue that the disorder is
rather large in the modified surface layer.

In the MgB$_2$/Ag-spot junctions, due to the very large apparent
contact area (the Ag spots have $\varnothing=150-250\mu$m, as can
be seen from Fig.~1), we probably measure an effective average of
the properties of many crystals with different orientation, of
their surfaces and also of the intergrain material, possibly
having different composition. Therefore, the directionality of the
contact is completely lost. Moreover, the absence of pressure in
the contact region makes it impossible to remove the probable
modified surface layer, which is therefore expected to play a
major role in these junctions (proved by the very large $\Gamma$
values) and we can argue that we are actually measuring the
Andreev reflection properties of MgB$_2$ surfaces in the
\emph{dirty} limit. In this conditions the paper of Liu \emph{et
al.} \cite{ref22} predicts the presence of a single gap with a
low-temperature value $\sim 0.6 \Delta_\mathrm{BCS}$, a reduced
$T_\mathrm{c}$ and a BCS-like temperature dependency. The results
shown in fig. 4 (b) seem quite compatible with these predictions.

The previous argument can also help to interpret the results of
many other groups who observed, together with $\Delta$ values
similar to that we measured in the dirty limit, Andreev reflection
features with rather small amplitude \cite{ref7,ref8} or largely
broadened STM tunneling curves \cite{ref5}. Incidentally, if we
simulate the tunneling curves that can be obtained by using
$\Delta=4.9$~meV and $\Gamma\approx 3$~meV (approximately the
values we obtained in MgB$\ped{2}$/Ag spot junctions) and $Z \sim
5$, we get d$I$/d$V$ curves very similar to those shown in
\cite{ref5}.

Of course, the previous discussion opens some questions concerning
the role of impurities in MgB$_2$. Contrary to what is expected
for a superconductor in pure \emph{s}-wave symmetry, here the
nonmagnetic impurities seem to have a big effect on the
superconducting properties of the material.

In conclusion, even if a complete temperature and magnetic field
dependency of the small out-of-plane 3D gap and the independent
observation of the large in-plane quasi-2D gap still have to be
done, we suggest that the Andreev reflection results obtained in
high-quality single-crystal-like samples are compatible with the
two-gap model of superconductivity in MgB$_2$ \cite{ref22}, with
the presence of a modified layer at the surface of the crystals
and with an important and non-conventional role of the impurities
in this material.

EDISON acknowledges the Lecco Laboratories of the CNR-TeMPE for
making its facilities available for material preparation. One of
the authors (V.A.S.) acknowledges the partial support by the
Russian Foundation for Basic Research (grant No 99-02-17877) and
by the Russian Ministry of Science and Technical Policy within the
program ``Actual Problems of Condensed Matter Physics'' (grant No
96001).

\end{document}